\documentclass[aps,prl,preprint,superscriptaddress]{revtex4-1}
\usepackage{graphicx}
\usepackage{dcolumn}
\usepackage{bm}
\usepackage[latin1]{inputenc}
\begin{document}


\title{Optical beam shifts in graphene and single-layer Boron-Nitride}


\author{Michele Merano}
\email[]{michele.merano@unipd.it}

\affiliation{Dipartimento di Fisica e Astronomia G. Galilei, Universit$\grave{a}$ degli studi di Padova, via Marzolo 8, 35131 Padova, Italy}


\date{\today}

\begin{abstract}
Optical beam shifts from a free-standing two-dimensional atomic crystal are investigated. In contrast to a three-dimensional crystal the magnitude of the Goos-H$\rm \ddot{a}$nchen shift depends on the surface susceptibility of the crystal and not on the wavelength of the incident light beam. The surface conductivity of the atomically thin crystal is less important in this context because it enters in the expression of the shifts only as a second order parameter.  In analogy to a three-dimensional crystal the magnitudes of the Imbert-Fedorov shift and of the angular shifts depend respectively on the wavelength and on the square of the beam angular aperture. 
\end{abstract}

\maketitle

The Goos-H$\rm \ddot{a}$nchen (GH) shift \cite{Goos47, Bliokh13, Merano07} is a spatial displacement, from the prediction of geometrical optics, of a light beam reflected at an interface. This translation occurs in the plane of incidence perpendicularly to the direction of propagation of the light beam. In a ray model description, that was first employed by Newton \cite{Newton} and then adopted by the optical beam shift community \cite{Anicin78, Carniglia77, Jackson}, the GH effect seems to define an effective depth for reflection. The lateral displacement of the beam, can be explained by considering a plane (the equivalent plane of reflection \cite{Anicin78}), placed to a certain distance beyond or before the boundary, that effectively reflects the light (Fig. 1). This phenomenon, when interfaces between two homogeneous media of different optical properties are considered, is of the order of the wavelength of the incident light showing that it is essentially a diffractive correction to the laws of geometrical optics. 

Other diffractive effects affect the reflection of a light beam. The Imbert-Fedorov (IF) shift \cite{Imbert72, Fedorov55, Bliokh13, Bliokh06, Bliokh07} is a spatial displacement of the reflected beam that occurs orthogonally both to the plane of incidence and to the propagation direction. The GH and the IF effects depend on the polarization of the incident beam. While the eigenmodes of the GH shift are $p$ and $s$ linearly polarized modes, the eigenmodes of the IF effect are circularly polarized waves \cite{Bliokh13}. Spatial shifts are not the only corrections to geometrical reflection for a light beam incident at an interface. Angular deviation of the beam axis from the ray optics predictions have been reported as well \cite{Merano09, Merano102}. Optical beam shifts are not limited to reflection but they are observable also in transmitted light beams \cite{Bliokh13, Bliokh06, Bliokh07}.

These phenomena can be highly dependent on the optical properties of the interface under investigation. A spatial GH shift is observed in total internal reflection while the angular GH (AGH) shift is observed in external reflection from an insulating transparent dielectric \cite{Merano09, Merano102}. The GH shift in total internal reflection is positive \cite{Jackson} but it is negative for $p$ polarized light in metallic reflection \cite{Merano07}. Interfaces with losses induce at the same time the GH and the AGH effect \cite{Aiello092}. When reflection from multi-layers or gratings \cite{Yang14} or metamaterials \cite{Kivshar03, Yin13, Xu15, Luo15, Ling15, Liu15, Shaltout15, Lee15, Gopal16} is considered all these optical beam shifts can be greatly enhanced.

In 2004 two-dimensional (2D) crystals \cite{Novoselov2004, Novoselov2005}, a completely new class of optical interfaces, were discovered. They are single atomic planes that are stable under ambient conditions and are continuous on a macroscopic scale. The family of experimentally isolated 2D crystals broadens each year and there are now tens of such materials. This derives into a rich variety of electronic properties including metals, semimetals, insulators and semiconductors with direct and indirect band gaps ranging from ultraviolet to infrared throughout the visible range \cite{Heine14}.         

\begin{figure}
\includegraphics{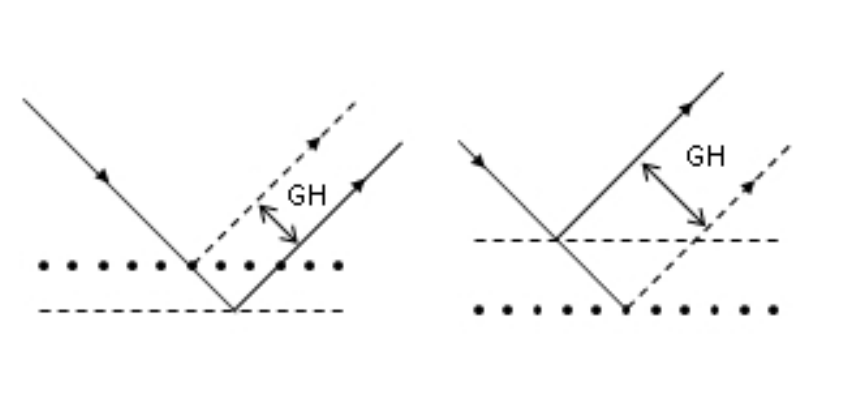}
\caption{\label{EPR} A beam of light is incident on a 2D atomic crystal. The dots indicate the crystal plane; the horizontal dash line indicates the equivalent plane of reflection. For a positive Goos-H$\rm \ddot{a}$nchen shift the equivalent plane of reflection is placed beyond the crystal plane (image on the left), for a negative shift it is placed before it (image on the right).}
\end{figure}

Also the optical properties of 2D crystals confirm their macroscopic character. Their optical response depends on their macroscopic surface susceptibility ($\chi$) and surface conductivity ($\sigma$) \cite{Luca16}. In analogy to a bulk crystal these macroscopic quantities can be conceptually introduced without resorting to a microscopic atomic description. These parameters are experimentally accessible via ellipsometry \cite{Merano16}. Other experimental techniques like optical contrast, reflectometry or absorption can be useful to extract the values of $\chi$ and $\sigma$ \cite{Merano16}.

\begin{figure}
\includegraphics{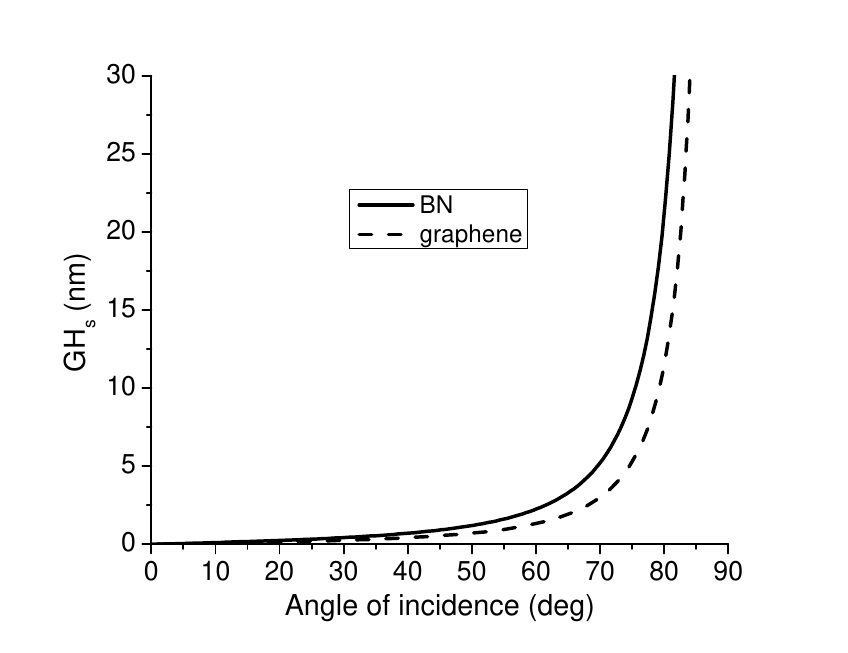}
\caption{\label{} Goos-H$\rm \ddot{a}$nchen shift for a monochromatic ($\lambda=633$ nm), $s$ polarized Gaussian beam incident on a free-standing two-dimensional crystal suspended in vacuum. The magnitude of the effect is of the order of the crystal surface susceptibility, and for geometrical reasons it is greatly enhanced at grazing incidence.}
\end{figure}

The purpose of this paper is to study optical beam shifts when a monochromatic Gaussian light beam is reflected from a single-layer 2D crystal. Theoretical and experimental studies of optical beam shifts from graphene have been carried out recently \cite{Ornigotti15, Hermosa16, Cheng14, Li14, Jalil11, Wen16}. They have all concerned the case of graphene deposited on a 3D substrate. These works considered specific cases where the GH shift from the substrate is enhanced by the presence of graphene.

Here a different approach is adopted. In order to grasp the basic role of a 2D material in optical beam shifts the focus is mainly on a generic free-standing atomic crystal. In particular I address the following questions: is the equivalent plane of reflection for the GH effect placed within a distance of the order of the wavelength from the atomic crystal, exactly as what happens for a 3D crystal? Or should another scale length be considered? And in the case of the IF effect, what to expect? What are the differences in between insulating and conducting materials or, in other terms, what is the role of $\chi$ and $\sigma$ in this context?

The polarization-dependent displacements parallel and perpendicular to the plane of incidence for a Gaussian light beam reflected from a planar interface, have been deduced in ref. \cite{Aiello08}, where the paraxial approximation corrected up to first-order derivatives has been used. Formulae 4 and 5 of ref. \cite{Aiello08} provide the spatial and angular GH and IF shifts once the reflection coefficients for the interface are known. For a general 2D crystal, Fresnel coefficients were derived in ref. \cite{Merano16}. Here I report those for a free-standing crystal suspended in vacuum
\begin{eqnarray}
r_{s}=-\frac{ik\chi+\sigma\eta}{ik\chi+\sigma\eta +2\cos\theta};\quad  r_{p}=\frac{(ik\chi+\sigma\eta)\cos\theta}{(ik\chi+\sigma\eta)\cos\theta+2}\quad
\end{eqnarray}
where $\theta$ is the angle of incidence, $k=2\pi / \lambda$ is the vacuum wavector and $\eta$ is the impedance of vacuum. 

\begin{figure}
\includegraphics{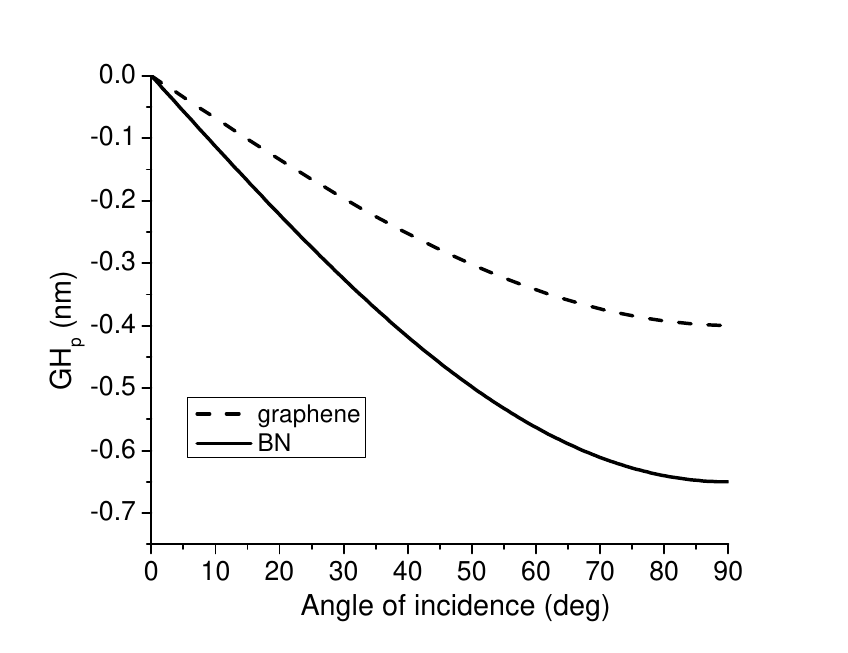}
\caption{\label{} Goos-H$\rm \ddot{a}$nchen shift for a monochromatic ($\lambda=633$ nm), $p$ polarized Gaussian beam incident on a free-standing two-dimensional crystal suspended in vacuum. The magnitude of the effect is of the order of the crystal surface susceptibility.}
\end{figure}

Reference \cite{Merano16} shows also how to extract from experimental data both $\chi$ and $\sigma$. For the purposes of this paper conducting graphene and insulating BN will be considered. Assuming an incident wavelength $\lambda=$ 633 nm, for graphene $\sigma=6.08\cdot 10^{-5}\pm 2\cdot10^{-5} \Omega^{-1}$ and $\chi=8\cdot 10^{-10}\pm 3\cdot 10^{-10}$ m \cite{Merano16}. At the same $\lambda$, from ref. \cite{Blake2011} that reports optical contrast measurements of single layer BN on top of a $\rm SiO_{2}/Si$ wafer with a $\rm SiO_{2}$ thickness of 290 nm, ref. \cite{Merano16sw} extracts a value of $\chi = (1.3 \pm 0.1) \cdot 10^{-9} $ m and an upper limit for $\sigma \leq 2\cdot 10^{-6} \ \Omega^{-1}$. In the following for a single-layer BN $\sigma= 0 \Omega^{-1}$ will be assumed.

As the result of a straightforward calculation one obtains for the GH shift of an $s$ and a $p$ polarized beam respectively  
\begin{eqnarray}
\label{conductingGH}
GH_s(\theta)=\frac{2\chi \sin\theta}{4 \cos^2\theta+k^2\chi^2+\sigma^2\eta^2+4 \sigma \eta \cos \theta}
 \\
GH_p(\theta)=-\frac{2\chi \sin\theta}{4+(k^2\chi^2 +\sigma^2\eta^2)\cos^2\theta+4 \sigma \eta \cos \theta}
\end{eqnarray}
From these formulae it is clear that the GH shift has the dimension of a distance. Surprisingly the magnitude of the effect does not depend primarily on $\lambda$ but on $\chi$, which also fixes its sign. The role of $\sigma$ is that of reducing its absolute value. In the visible spectral range, the GH shift is positive for an $s$ polarized beam and negative for a $p$ polarized beam, both for conducting graphene and for insulating BN. (The sign convention adopted in ref. \cite{Aiello08} is the opposite of the one usually adopted in literature \cite{Merano07}. Here I follow this last one.)  

Figures 2 and 3 show respectively the GH shift for an $s$ and a $p$ polarized beam at $\lambda= 633$ nm, for the two atomic crystals here considered. The absolute value of both shifts is greater for BN than for graphene because of their different value for $\chi$. At grazing incidence, shifts for both polarizations do not diverge.

\newpage
I consider now the spatial IF effect. For the circular plus/minus ($\sigma \pm$) polarization basis I obtain:
\begin{eqnarray}
\label{conductingIF}
IF_{\sigma \pm} (\theta)=\mp \frac{\cot\theta}{k}\bigg(1+ \qquad \qquad \qquad \qquad \qquad \qquad \qquad \\
\frac{\cos^2\theta(4+k^2 \chi^2+\sigma^2\eta^2)+\sigma \eta \cos \theta (3+\cos 2\theta)}{2(1+\cos^4 \theta+\sigma \eta \cos\theta(1+\cos^2\theta))+(k^2 \chi^2+\sigma^2\eta^2)\cos^2\theta}\bigg) \nonumber
\end{eqnarray}
Remarkably the magnitude of this effect depends primarily on $\lambda$ as what happens for 3D crystals \cite{Bliokh06, Bliokh07}. Figure 4 reports the IF effect for a $\sigma +$ polarized beam at $\lambda= 633$ nm. The different optical properties of graphene and BN are almost irrelevant here contrary to what happens for the GH effect. This different behavior of the GH and the IF shifts for a 2D crystal is the main result of this paper.

Due to beam propagation, angular effects may arise. In particular the AGH effect for an $s$ and a $p$ polarized beam is given by 
\begin{eqnarray}
\label{cAGH}
AGH_s(\theta)=\theta^2_0\frac{2(\sigma \eta+2\cos\theta) \sin\theta}{4 \cos^2\theta+k^2\chi^2+\sigma^2\eta^2+4 \sigma \eta \cos \theta} \\
AGH_p(\theta)=-\theta^2_0\frac{2 (2+\sigma \eta \cos\theta) \tan \theta}{4+(k^2\chi^2 +\sigma^2\eta^2)\cos^2\theta+4 \sigma \eta \cos \theta}
\end{eqnarray}
The AGH depends on the square of the beam angular aperture ($\theta_0$). This is the same behavior observed in reflection from a 3D crystal \cite{Merano09, Merano102}. Polarizations $p$ and $s$ differ in the sign of the effect. Insulating and conducting materials differ mainly for the value of the $s$ polarized shift at grazing incidence. For insulators it tends to zero while for a conductor it tends to a positive value (Fig.5). In the approximation here considered, the $p$ polarized shift diverges at grazing incidence.

\begin{figure}
\includegraphics{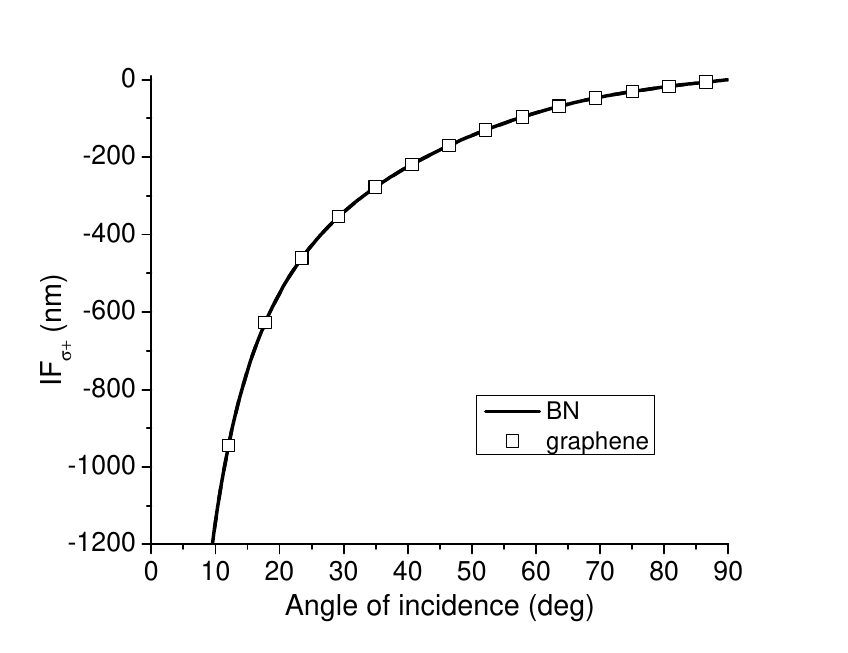}
\caption{\label{} Imbert-Fedorov shift for a monochromatic ($\lambda=633$ nm), circular plus polarized Gaussian beam incident on a free-standing two-dimensional crystal suspended in vacuum. The magnitude of the effect is of the order of the incident wavelength.}
\end{figure}

It is not difficult to extend the analysis here done, based on the Fresnel coefficients of ref. \cite{Merano16}, to a 2D crystal deposited on a substrate. Results published in ref. \cite{Ornigotti15} for a graphene-coated glass, both for the cases of internal and external reflection, are consistent in magnitude with those obtained here. The exception is in proximity of the Brewster's angle and the critical angle, where second-order derivatives effects may become relevant and equations in ref. \cite{Aiello08} are not any more valid. Experimentally a giant spatial GH effect in graphene was observed in a total internal configuration \cite{Li14} scheme. The shift presented in \cite{Li14} is orders of magnitude larger than that expected from the present results. Measurements in ref. \cite{Li14} were done with a focused beam but the angular shift contribution was not accounted for. The theoretical model of ref. \cite{Hermosa16} claims to agree with the order of magnitude of this (angular shift included) GH shift  \cite{Li14}. However, in ref. \cite{Hermosa16} Artmann's formula is mentioned to give a spatial GH shift of the order of the nanometer which is similar to that shown in the present work. 

Graphene transmits just less than 98\% of the incident light (BN even more) \cite{Merano16}. This brings in a practical difficulty to observe a GH shift from graphene, as most of the reflected light detected would be from the substrate. A possible way to observe it is by embedding graphene in a dielectric medium. The light reflected from graphene in this case should be sufficient to perform a measurement in reflection.    

I have studied optical beam shifts in reflection from a free-standing 2D atomic crystal suspended in vacuum. The GH shift is of the order of $\chi$ while the IF shift is of the order of $\lambda$. The different behavior of the two effects is a hint of their different nature. Transverse shifts are responsible for the conservation of the total angular momentum of an electromagnetic beam including the spin part \cite{Bliokh06, Bliokh07}. In a ray optics interpretation \cite{Anicin78, Carniglia77, Newton} the longitudinal shift defines an equivalent plane of reflection. For a 3D crystal this plane is placed at a distance of the order of $\lambda$ from the physical interface. This would be an enormous distance for an atomically thin material if compared to the typical inter-layer spacing of the original bulk material. The surface susceptibility of these materials instead is of the order one nanometer only \cite{Merano16, Merano16sw} showing the consistency of the ray optics interpretation of the GH effect \cite{Anicin78, Carniglia77, Newton} even for atomically thin interfaces.  

\begin{figure}
\includegraphics{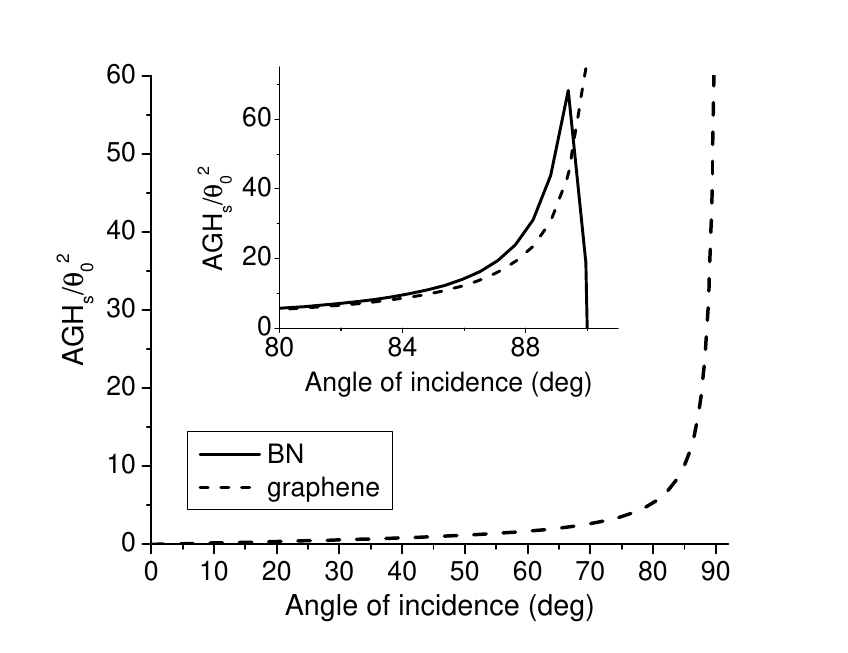}
\caption{\label{} Angular Goos-H$\rm \ddot{a}$nchen shift for a monochromatic ($\lambda=633$ nm), $s$ polarized Gaussian beam incident on a free-standing two-dimensional crystal suspended in vacuum. The magnitude of the effect is of the order of the square of the beam angular aperture. Inset: difference in between conducting graphene and insulating BN.}
\end{figure}

The different behavior of the GH shift from single-layer 2D crystals to 3D crystals raises important questions. How does the optical transition from an atomic crystal to a bulk crystal occur? How many atomic layers we need to observe an optical response typical of a bulk material? How do hetero-structures behave in this context? The results here reported, raising important new questions for future research, show once more how much exciting are the optical properties of 2D crystals.

\bibliography{letter}

\end{document}